\documentclass[conference,a4paper]{IEEEtran}

\ifCLASSINFOpdf
\else
\fi
\usepackage[left=1.57cm,right=1.57cm,top=0.95cm,bottom=2.54cm]{geometry}
\usepackage{listings}
\usepackage{setspace}
\usepackage{graphicx}
\usepackage{tabularx,booktabs}
\usepackage{dingbat}
\usepackage{diagbox}
\usepackage{multirow} 
\usepackage[hyphens]{url}
\usepackage[hidelinks,breaklinks,bookmarks=false]{hyperref}
\usepackage{xcolor}
\usepackage{amsfonts, amsmath, amsthm, amssymb}
\usepackage{subcaption}
\hypersetup{breaklinks=true}
\usepackage{tikz}
\usepackage{textcomp}
\usepackage{lipsum}
\usepackage{svg}
\IEEEoverridecommandlockouts

\usepackage{graphicx}
\graphicspath{{figures/}}
\begin{document}
\bstctlcite{IEEEexample:BSTcontrol}
%
\title{Streamable Neural Video Compression: A Mixed Precision Approach for Cross-Platform Deployment}

\author{
    \IEEEauthorblockN{Kasidis Arunruangsirilert\IEEEauthorrefmark{1}, Heming Sun\IEEEauthorrefmark{2}, Jiro Katto\IEEEauthorrefmark{1}}
    \IEEEauthorblockA{\IEEEauthorrefmark{1}Department of Computer Science and Communications Engineering, Waseda University, Tokyo, Japan}
    \IEEEauthorblockA{\IEEEauthorrefmark{2}School of Computing, Institute of Science Tokyo, Tokyo, Japan
    \\\{kasidis, katto\}@katto.comm.waseda.ac.jp, son.k.2b4a@m.isct.ac.jp}
}
%

\maketitle

\setstretch{0.90}
\begin{abstract}

Neural Video Codecs (NVCs) offer unprecedented rate-distortion performance, making them highly attractive for bandwidth-constrained environments like 5G cellular networks and emerging satellite direct-to-cell (D2C) links. However, deploying NVCs in real-world streaming applications is severely hindered by cross-platform floating-point non-determinism, which causes arithmetic entropy coders to desynchronize and crash across different GPU architectures. While recent integer-based quantization methods address this, they incur either massive degradation in compression efficiency (INT8) or severe computational bottlenecks by bypassing hardware acceleration (INT16). In this paper, we propose a streamable, client-server NVC architecture featuring a novel Mixed Precision (FP16/FP32) strategy. By strategically executing P-frames in hardware-accelerated FP16 for real-time throughput, while forcing I-frames and periodic feature-adapter resets to IEEE-754 compliant FP32, we keep the computations at these critical boundaries consistent across GPU die variants. Through extensive cross-encode/decode evaluations across 12 GPUs spanning four architectural generations, we demonstrate that our approach largely eliminates intra-generation fragmentation and substantially broadens cross-die interoperability, achieving seamless cross-generation decodability for recent architectures at 1080p. Crucially, this is achieved with a negligible impact on compression efficiency. Furthermore, we evaluate the system's end-to-end latency across diverse real-world networks, including Wi-Fi 6, 5G NR (FDD/TDD), and Starlink D2C, proving the practical viability of streamable learned video compression while highlighting unique challenges in Non-Terrestrial Networks.

\end{abstract}

\begin{IEEEkeywords}
Cross-Platform Interoperability, GPU Architecture, Learned Video Compression, Mixed Precision, Real-Time Streaming
\end{IEEEkeywords}


\setstretch{0.890}

%
\IEEEpeerreviewmaketitle

\vspace{-3mm}
\section{Introduction}

Neural Video Codecs (NVCs) have recently demonstrated remarkable rate-distortion (RD) performance, increasingly surpassing the coding efficiency of state-of-the-art traditional video coding standards such as High Efficiency Video Coding (HEVC) \cite{6316136} and Versatile Video Coding (VVC) \cite{9503377, 10.1145/3746027.3755598, 11095025}. This superior compression capability is particularly vital for low-throughput network environments, ranging from highly congested 5G New Radio (NR) cellular networks in dense urban areas \cite{11133825} to emerging Non-Terrestrial Networks (NTNs) such as Starlink Direct-to-Cell (D2C) satellite services \cite{11099929}. Concurrently, the explosion of User-Generated Content (UGC) and the rapid evolution of Generative AI have fundamentally shifted video traffic patterns \cite{4801529, 10731639, 11016906}. Users and AI agents now require the ability to create, transmit, and consume high-resolution video in real time, moving beyond the traditional Video-on-Demand (VoD) paradigm \cite{10089871}. Consequently, the development of streamable, real-time NVCs has become one of the most promising ways to overcome network bandwidth constraints. 

Despite their theoretical performance advantages, a major drawback preventing the widespread deployment of learned video compression in real-world applications is cross-platform bitstream compatibility. NVCs inherently rely on floating-point arithmetic (FP16 or FP32) for deep neural network inference. However, floating-point operations often yield non-deterministic results across different hardware due to variation in GPU architectures, Streaming Multiprocessor (SM) counts, and backend algorithm selections. In a standard Learned Video Compression (LVC) architecture, even a single-bit numerical divergence in the reconstructed latent features can cause the arithmetic entropy coder to desynchronize between the encoder and the decoder, leading to catastrophic error propagation, corrupted video outputs, or, in extreme cases, crashing the decoder altogether \cite{10743422, 10.1145/3581783.3611955}.

\begin{figure}[t!]
\centering\includesvg[width=0.73\linewidth,inkscapelatex=false]{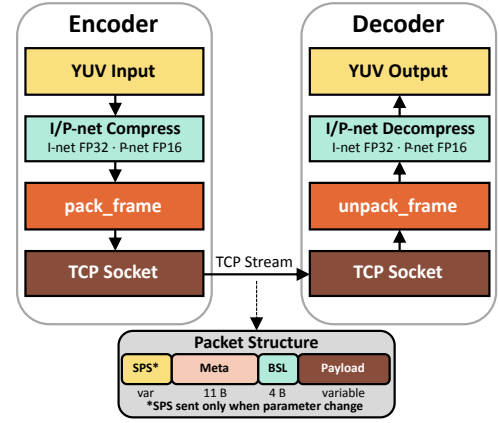}
\caption{System Architecture of the Proposed Method}
\label{fig:SystemArchitecture}
\vspace{-7mm}
\end{figure}

To address floating-point non-determinism, several recent works have proposed integer-based quantization or calibration bitstreams \cite{10448359, tian2023effortlesscrossplatformvideocodec}. Notably, recent studies, including the integer-centric architectures evaluated in \cite{jia2026integercentric}, advocate for converting floating-point NVCs entirely to 8-bit (INT8) or 16-bit (INT16) integers. However, these integer approaches introduce severe practical trade-offs. INT8 quantization is often too coarse for the delicate probability distributions required by entropy models, resulting in massive degradation in compression efficiency, e.g., up to an 80\% BD-Rate penalty in models like MobileNVC \cite{10484439, jia2026integercentric}. Conversely, while using INT16 and INT32 can resolve floating-point determinism issues and maintain the compression efficiency, they are rarely supported by efficient hardware acceleration on standard deployment platforms. Modern AI accelerators, including NVIDIA Tensor Cores and Neural Processing Units (NPUs) on modern chipsets, are overwhelmingly optimized for FP16 and INT8 matrix multiplications \cite{burns_2024, nvidia_tensorcore}. Using INT16 on this hardware forces it to cast INT16 to FP32, perform the calculation, and cast the result back to INT16, incurring massive computational overhead. For example, in DCVC-RT, this results in up to a 5× slowdown in coding speed compared to FP16 \cite{11095025}. Therefore, successfully leveraging FP16 and FP32 operations is crucial to achieving the real-time processing speeds required for high-resolution video, as the hardware acceleration already exists.

In this paper, we present a streamable, cross-platform NVC architecture that bridges the gap between cross-platform decodability and real-time floating-point acceleration. Building upon the state-of-the-art DCVC-RT framework \cite{11095025}, we introduce a lightweight streaming encapsulation format alongside a novel Mixed Precision (FP16/FP32) strategy. Through a systematic investigation of cross-GPU decodability, we reveal that floating-point compatibility is governed by a complex hierarchy tied to hardware architecture, die similarity, and video resolution. While standard FP16 execution severely restricts inter-decodability to highly similar GPU dies, our Mixed Precision and pure FP32 approaches significantly broaden this compatibility across multiple modern GPU generations, navigating complex hardware and resolution boundaries. Crucially, we demonstrate that these precision modes achieve robust cross-die determinism with a negligible impact on compression efficiency, successfully bypassing the severe computational overhead and quality penalties of integer quantization.

\begin{itemize}
    \item We transform the offline DCVC-RT framework into a deployable, client-server streaming architecture. To resolve hardware decodability crashes without sacrificing real-time hardware acceleration, we introduce a novel Mixed Precision (FP16/FP32) operational mode.
    \item We systematically evaluate bitstream inter-decodability and compression efficiency across 12 distinct GPUs from four architectural generations. We map the complex boundaries of hardware-induced floating-point non-determinism, demonstrating where Mixed Precision successfully unifies intra-generation compatibility and where resolution-dependent fractures remain.
    \item We identify and address architectural barriers in split encoder-decoder deployments by introducing a deterministic feature-adapter refresh mechanism and enforcing Decoded Picture Buffer (DPB) state synchronization.
    \item We deploy and evaluate the end-to-end latency of our streamable NVC across diverse operational networks, including Ethernet, Wi-Fi 6, terrestrial cellular (LTE/5G NR), and Starlink D2C, providing practical insights into deployment over both stable and highly volatile links.
\end{itemize}

\vspace{-1.5mm}

\section{Proposed Method}
\vspace{-1mm}
\subsection{Background}

The challenge of cross-platform decodability in NVCs stems from the dual-execution-unit architecture of modern NVIDIA GPUs. General-purpose CUDA cores provide IEEE-754 compliant FP32 arithmetic with correctly rounded operations, although the library still chooses their order. In contrast, specialized Tensor Cores deliver massive throughput gains by operating on lower-precision FP16 inputs but do not guarantee bit-exact results across different GPU models. In a typical streaming scenario, where a high-end encoder GPU serves a lower-tier consumer decoder GPU, these minor numerical divergences become critical. If the probability distributions generated by the two GPUs differ even slightly, the arithmetic entropy coder desynchronizes, leading to catastrophic decoding failure. Within a generation, the primary source of this non-determinism is algorithm selection that depends on the GPU die variant. Within a single GPU generation (e.g., Blackwell), NVIDIA manufactures multiple dies with varying Streaming Multiprocessor (SM) counts. The cuDNN library dynamically selects different internal Tensor Core algorithms based on the available SMs to optimize performance. When two GPUs from the same generation but with different dies execute the same operation, these distinct algorithms produce outputs that differ in their least significant bits. Any such difference in the decoder-side computations alters the entropy model's probability tables, causing the decoder to crash.

\vspace{-1mm}

\subsection{Streaming Codec Architecture}

The reference DCVC-RT implementation is a single-process offline tool: the encoder and decoder share the same program, the same model weights in memory, and the same GPU. Frames are compressed one at a time, and the results are stored in a memory buffer, with quality metrics computed after the entire sequence has been processed. Hence, this design cannot be used for streaming. We restructure DCVC-RT as a client-server streaming system. The encoder runs as a TCP server, accepting a connection from a remote decoder and transmitting each compressed frame immediately as it is produced, without waiting for the end of the GOP. TCP was chosen to guarantee reliable delivery, as NVCs exhibit extreme sensitivity to data corruption and lack error-resilience mechanisms. The decoder connects as a TCP client, receives incoming frames, decompresses them, and outputs the result (see Fig. \ref{fig:SystemArchitecture}). This design minimizes the latency to the time required to compress and decompress the frame plus network transit time, as no frame buffering or reordering is performed.

\begin{table}[!tbp]

\setstretch{0.75}
\caption{Per-Frame Packet Structure}
\vspace{-2mm}
\centering
\label{tab:packet}
\resizebox{8.5cm}{!}{\begin{tabular}{@{}lll@{}}
\toprule
Field                 & Size & Description  \\\midrule
Length Prefix & 4 bytes & Total packet length for TCP demarcation\\
SPS Length & 4 bytes & Byte length of the SPS payload; zero if unchanged\\
SPS Payload & Variable & Sequence parameter set\\
Frame Type & 1 byte & 0 = I-frame, 1 = P-frame\\
Frame Index & 4 bytes & Sequential frame counter \\
SPS Identifier & 1 byte & Reference to the active parameter set \\
Quantization Parameter & 1 byte & Per-frame QP value \\
Bitstream Length & 4 bytes & Byte length of the entropy-coded payload \\
Entropy-Coded Bitstream & Variable & Raw arithmetic-coded output, same as DCVC-RT\\
\bottomrule
\end{tabular}}
\vspace{-7mm}
\end{table}

To enable both real-time streaming and offline batch evaluation, we define a unified binary packet format that is identical whether sent over a TCP connection or written to a file on disk. Any binary file produced by the offline encoder can be replayed as a live stream, and any live stream can be recorded to a file for later analysis. The packet structure is shown in Table \ref{tab:packet}. Each compressed frame is wrapped in a self-contained packet consisting of a conditional sequence parameter set (SPS) carrying frame dimensions and coding flags, a frame type and index, a quantization parameter, and the raw entropy-coded bitstream produced by DCVC-RT's arithmetic coder without modification. An outer 4-byte length prefix allows a TCP receiver to read exactly one complete packet with two network calls, regardless of frame size or network fragmentation. The SPS is embedded in the packet only when it changes, typically once at the start of a sequence. Subsequent packets carry only a one-byte SPS identifier, keeping per-frame overhead minimal. This design allows the decoder to configure itself from the bitstream alone without any separate control channel.

\vspace{-1mm}

\subsection{Mixed Precision Implementation}

\begin{figure*}[t!]
\centering\includesvg[width=0.87\linewidth,inkscapelatex=false]{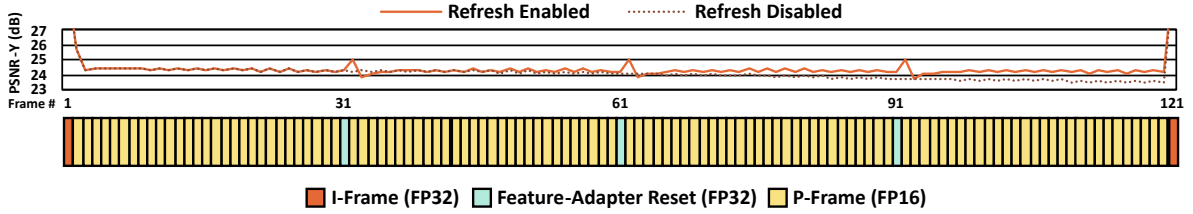}
\caption{PSNR progression over a 120-frame GOP with and without periodic feature-adapter refresh. Applying deterministic FP32 resets every 30 frames effectively bounds accumulated error drift and prevents quality degradation.}
\label{fig:GOPView}
\vspace{-7mm}
\end{figure*}

The DCVC-RT model consists of two neural networks: I-net (DMCI) and P-net (DMC), which are used for I-frame and P-frame coding, respectively. In the reference implementation, both models are executed entirely in half-precision (FP16), prioritizing coding throughput at the cost of compatibility across different GPU architectures. Conversely, running the whole pipeline in FP32 proves extremely computationally expensive (see Section \ref{sec:coding_thpt}). To resolve this, we introduce a \textit{Mixed Precision} mode that leverages the asymmetric roles of frames within a GOP (see Fig. \ref{fig:GOPView}). 

In our implementation, the I-net operates in FP32, while the P-net operates in FP16; both directly reuse the original DCVC-RT pretrained weights with no fine-tuning or recalibration, with precision applied by casting each network's weights and activations at model load time. The P-frame network processes the vast majority of the workload. Our bitstream hashes (Section \ref{sec:cross_dec}) show that its FP16 computations are bit-identical across the die variants of a generation in most configurations, so the P-frames can remain on the Tensor Cores. The die-dependent computations instead arise in the I-frame path. The I-frame encodes the base reference picture that seeds the DPB for the entire GOP, so any difference in its output corrupts the context for all subsequent P-frames. By executing the I-frame network in FP32, where the die variants of a generation agree, we remove this source of mismatch at every GOP boundary. Because the I-frame network executes only once per GOP, the computational cost remains minimal.

\vspace{-1mm}

\subsection{Deterministic Feature-Adapter Reset}

DCVC-RT's P-net maintains a propagated feature tensor in a Decoded Picture Buffer (DPB) that carries temporal context from one frame to the next. Over a long sequence, small numerical errors can accumulate in this tensor, resulting in visible artifacts. To prevent quality degradation from unchecked drift, DCVC-RT provides a periodic feature-adapter reset every \textit{reset\_interval} P-frames (30 by default; see Fig. \ref{fig:GOPView}), in which the synthesis network re-derives a pixel-domain reference from the current frame state and stores it as a fresh anchor. Compared to a full I-frame, the computational cost is minimal, as this is signaled to the decoder by a single bit in the packet header and executes in a few milliseconds. In the original implementation, this reset runs at FP16 via tensor cores. When the encoder and decoder run on different GPU die variants, the tensor core outputs differ in their least significant bits, and the hard clamp to [0, 1] applied immediately afterward acts as a nonlinear amplifier: values that land on different sides of the clamp boundary on two different GPUs produce a discrete, non-vanishing difference that propagates through subsequent feature processing into the arithmetic coder. Once the probability tables disagree between the encoder and decoder, decoding fails. 

We mitigate this by running the reset step in FP32 on CUDA cores regardless of the model's operating precision. At model load time, we cache a dedicated FP32 copy of the synthesis network. At each reset point, this cached copy is invoked through PyTorch's standard code path, bypassing the custom CUDA kernel and its tensor core dispatch. By explicitly bypassing customized Tensor Core hardware kernels and forcing the PyTorch-native FP32 execution path, the synthesis and clamping operations occur consistently across GPU die variants. The result is then cast back to the native precision. This keeps the anchor refresh consistent across GPU die variants.

\vspace{-1mm}

\subsection{DPB State Mismatch Fix for Split Deployments}

A further challenge arises from splitting the encoder and decoder across separate processes. In the reference implementation, the encoder and decoder share one DPB object, so any state written by the encoder's compress operation is immediately visible to the decoder's decompress operation within the same call stack. When separated across machines, each side maintains its own independent DPB, and any difference in the state will cause the two sides to diverge.

We identify such an inconsistency at feature-adapter reset boundaries. After compressing a P-frame, the encoder stores only the propagated feature in the DPB with no pixel reference, because the pixel reconstruction is not needed for subsequent compression. After decompressing the same P-frame, the decoder stores both the feature and the reconstructed pixel frame. When the next reset triggers, the encoder finds no pixel frame in the DPB and proceeds directly to compute a new one from the feature. The decoder, however, finds a pixel frame already present, and its reset guard evaluates differently, causing the two sides to take different internal code paths. The arithmetic coder's probability tables then diverge, and decoding fails at the very first reset point. To resolve this, we explicitly clear the decoder's pixel frame before resets to align its state with the encoder. We also implement a special case for frames immediately following an I-frame; because these frames lack the features required for a reset, both the encoder and decoder are forced to skip the reset procedure. This synchronized alignment ensures identical execution paths and maintains bitstream integrity throughout the streaming session.


\vspace{-1mm}

\subsection{Evaluation Setup and Methods}

\subsubsection{Test Dataset and GOP Format}

Following our previous works on video coding for live streaming \cite{11396901, 11417632}, we adopt the same GOP structure recommended by major streaming services, using a GOP length of two seconds \cite{google, twitch}. At the evaluated frame rate of 59.94 fps, this corresponds to approximately 120 frames per GOP. We set the feature-adapter reset interval to 30 frames, which divides evenly into the 120-frame GOP and ensures consistent performance characterization across all sequences. The reset triggers at frames 30, 60, and 90 within each GOP, producing four equal-length segments of bounded feature drift. We evaluate on the ITE Ultra-High Definition Standard Test Sequences (Series A) \cite{ITE_2016}. The dataset provides 10 sequences at 4K resolution (3840×2160) and 11 sequences at 8K resolution (7680×4320), all at 59.94 fps with a duration of 15 seconds per sequence. We use only the 10 4K-resolution sequences in this work. The dataset is in Standard Dynamic Range (SDR) with BT.2020 Wide Color Gamut (WCG). Since DCVC-RT operates in the BT.709 color space, all sequences were color-space converted to BT.709 Standard Color Gamut and written to raw YUV 4:2:0 planar files prior to evaluation.

\subsubsection{Coding Throughput}

\begin{table}[!tbp]
\setstretch{0.6}
\vspace{1.5mm}
\caption{Hardware and Software Configuration}
\vspace{-2mm}
\centering
\label{tab:hardware}
\resizebox{8.6cm}{!}{\begin{tabular}{@{}ll@{}}
\toprule
\multicolumn{2}{c}{\textbf{Workstation/Encoding Server}}\\
\midrule
Hardware                 & Description  \\\midrule
Processor (CPU) & Intel(R) Core(TM) Ultra 9 285K (8P/16E) (OC to 5.7 GHz) \\
Memory (RAM) & Dual-Channel DDR5 128 GB (4×32 GB) @ 5800 MT/s \\ 
Motherboard & ASUS TUF GAMING Z890-PRO WIFI \\
Storage (SSD) & WD\_BLACK SN850X 2000GB \\
\midrule
OS/Driver & Version \\\midrule
Operating System (OS) & Microsoft Windows Server 2025 Datacenter Build 26100\\
GeForce Driver & GeForce Game Ready Driver 581.80 (581.80)\\
Quadro/RTX PRO Driver& NVIDIA RTX Driver Release 580 R580 U5 (581.80)\\\midrule
\multicolumn{2}{c}{\textbf{Decoding Client}}\\
\midrule
Hardware                 & Description  \\\midrule
Chassis & Dell Pro Max 16 Plus MB16250 \\
Processor (CPU) & Intel(R) Core(TM) Ultra 9 285HX (8P/16E) (160W)\\
Memory (RAM) & Dual-Channel DDR5 128 GB CAMM2 @ 6400 MT/s \\ 
GPU & NVIDIA RTX PRO 5000 24GB Blackwell Laptop (170W)\\
Storage (SSD) & Dual 4TB PCIe 4.0 ×4 NVMe SSD (RAID 0) \\
Ethernet Adapter & Intel(R) Ethernet Controller I226-LM \\
Wi-Fi Adapter & Intel(R) Wi-Fi 7 BE200 320 MHz \\\midrule
OS/Driver & Version \\\midrule
Operating System (OS) & Microsoft Windows 11 Pro Build 26200\\
Driver& NVIDIA RTX Driver Release 580 R580 U5 (581.80)\\\midrule
Software Environment & Version \\\midrule
Python & 3.12.12 \\
PyTorch & 2.10.0+cu130 \\
NVIDIA CUDA Compiler & cuda\_13.1.r13.1\/compiler.36836380\_0 \\
C++ Build Tool & MSVC v143 VS 2022\\
FFmpeg & 2025-12-24-git-abb1524138\\

\bottomrule
\end{tabular}}
\vspace{-7mm}
\end{table}

Coding throughput experiments are conducted on the workstation detailed in Table \ref{tab:hardware}. To prevent hardware bottlenecks, GPUs are installed in the primary CPU-connected PCIe 5.0 ×16 slot, and source sequences are read from a high-performance NVMe SSD. Host-side processing variability is eliminated by pinning the Python process to CPU P-cores overclocked to 5.7 GHz under a High Performance power plan. GPU drivers remain constant, power limits are maximized, and custom CUDA kernels are recompiled for each specific GPU architecture. We evaluate the \textit{a05\_SteelPlant} sequence at QP 32 across four resolutions (720p, 1080p, 1440p, and 2160p) and three precision modes (FP16, Mixed Precision, and FP32). While the reference DCVC-RT evaluation \cite{11095025} isolates neural network inference time, we report true end-to-end wall-clock throughput (FPS). This metric encompasses all pipeline stages—disk I/O, host-to-device memory transfers, neural network inference, and packet framing—to accurately reflect real-world performance.

\subsubsection{Cross-Decodability}

To evaluate bitstream compatibility across heterogeneous GPU hardware, we decode the pre-encoded file \textit{a07\_River}, which contains significant motion and detail, at 1080p and 2160p using all available GPUs, under all three precision modes. The source bitstream was encoded at QP 47 to produce a high-fidelity operating point. For each combination of GPU, resolution, and precision mode, the Peak Signal-to-Noise Ratio (PSNR) of the decoded output is calculated against the input YUV source using the FFmpeg PSNR filter, and the luma component PSNR (PSNR-Y) is reported, consistent with standard practice in video coding evaluation \cite{11417632}. A result is classified as cross-decodable if the decoder completes without error and the PSNR-Y against the reference exceeds a threshold indicating proper reconstruction; a result is classified as incompatible if the decoder crashes or the PSNR-Y falls below this threshold due to arithmetic coder failure.

\subsubsection{Compression Efficiency}

To verify that the newly introduced precision mode does not degrade rate-distortion performance, we encode all ten ITE 4K sequences at seven QP values: 22, 27, 32, 37, 42, 47, and 52, under each of the three precision modes and two resolutions, and decode on the same GPU used for encoding. BD-Rate \cite{barman2024bjontegaarddeltabdtutorial} is computed using PSNR-Y against the uncompressed source as the distortion metric, evaluated sequence by sequence, and the average BD-Rate across all 10 sequences is reported as the summary statistic. BD-Rate differences between precision modes are computed pairwise with FP16 as the anchor.


\subsubsection{End-to-End Latency}

Since the physical separation required for outdoor cellular and satellite testing precludes true photon-to-photon measurement, we evaluate end-to-end Round-Trip Time (RTT) using solely the encoder's clock to eliminate synchronization errors. The encoder timestamps each frame before acquisition and calculates the elapsed time upon receiving a 4-byte frame-index acknowledgement (ACK) from the decoder over the TCP connection. This RTT encompasses the complete pipeline: disk read, host-to-device transfer, GPU encoding, network transit, GPU decoding, and ACK return. Baseline network latency is measured using 100 ICMP pings prior to each test. We evaluate performance across four network environments:

\vspace{-1mm}

\begin{itemize}
    \item \textbf{Wired Ethernet:} Connected via a residential FTTx.
    \item \textbf{Wi-Fi 6:} Connected via an Aruba AP-515 (802.11ax, 160 MHz channel bandwidth, same FTTx backhaul).
    \item \textbf{Cellular:} Utilizing a USB-tethered Snapdragon-equipped smartphone. We use Network Signal Guru to lock the User Equipment (UE) to specific configurations: FDD-LTE, TDD-LTE, FDD-NR, TDD-NR, and FDD-NR with Dynamic Spectrum Sharing (DSS).
    \item \textbf{LEO Satellite:} Utilizing Starlink Direct-to-Cell (D2C) via the same tethered smartphone setup.
\end{itemize}

\vspace{-1mm}

To ensure fair comparisons, the client laptop (Table \ref{tab:hardware}) operates on battery power across all tests, imposing a 55W GPU power limit. Under this power envelope, the decoder cannot sustain Mixed Precision's FP32 I-frame and reset passes in real time; we therefore evaluate in FP16, for which 720p (1280×720) at 30 fps is the highest stable operating point. Tests are conducted at QP 47 using the 60-second MINECRAFT sequence \cite{xiph.org}, retaining the 120-frame GOP and 30-frame refresh interval. \looseness=-5

\vspace{-1mm}

\section{Results and Analysis}

\begin{table*}[!tbp]
\setstretch{0.5}

\caption{Encoding/Decoding Throughput Across Various Configurations (FPS)}
\vspace{-2mm}
\centering
\label{tab:Thpt}
\resizebox{16.5cm}{!}{\begin{tabular}{@{}lllcccccccccccccccc@{}}
\toprule
\multirow{2.5}{*}{GPU}&\multirow{2.5}{*}{Generation}&\multirow{2.5}{*}{Die}& \multicolumn{3}{c}{720p/HD} & \multicolumn{3}{c}{1080p/FHD} & \multicolumn{3}{c}{1440p/QHD} & \multicolumn{3}{c}{2160p/4K} & \multicolumn{4}{c}{Mixed Perf. Penalty vs FP16 (\%)}\\
\cmidrule(lr){4-6} \cmidrule(lr){7-9} \cmidrule(lr){10-12} \cmidrule(lr){13-15} \cmidrule(lr){16-19}
&&&FP16&Mixed&FP32&FP16&Mixed&FP32&FP16&Mixed&FP32&FP16&Mixed&FP32&720p&1080p&1440p&2160p\\

\midrule
\multicolumn{19}{c}{\textbf{Encoding Throughput}}\\\midrule
RTX 2080 SUPER&Turing&TU104&68.9&65.7&26.6&32.4&30.8&12.2&18.3&17.5&6.9&8.3&7.9&3.1&4.64&4.94&4.37&4.82                      \\
RTX 3060 12 GB&Ampere&GA104&45.6&42.9&19.3&23.0&21.6&8.8&13.1&12.6&5.2&6.0&5.7&2.3&5.92&6.09&3.82&5.00                       \\
RTX 3080 12 GB&Ampere&GA102&98.0&90.9&41.3&46.6&43.4&18.6&28.7&27.3&11.3&13.0&12.4&5.1&7.24&6.87&4.88&4.62                   \\
RTX 3090&Ampere&GA102&106.0&99.1&45.8&51.5&48.2&21.6&30.4&29.0&13.1&14.6&14.0&5.7&6.51&6.41&4.61&4.11                        \\
RTX 4060 Ti 16 GB&Ada Lovelace&AD106&75.4&70.1&23.5&31.8&29.9&11.3&17.3&16.3&6.1&7.1&6.8&2.8&7.03&5.97&5.78&4.23             \\
RTX 4070 Ti&Ada Lovelace&AD104&120.5&111.0&42.2&53.4&50.1&20.5&29.2&27.6&11.0&12.1&11.7&4.8&7.88&6.18&5.48&3.31              \\
RTX 4070 Ti SUPER&Ada Lovelace&AD103&124.4&115.4&44.1&61.5&57.4&22.7&33.1&31.4&12.3&14.4&13.8&5.7&7.23&6.67&5.14&4.17        \\
RTX 6000 Ada Generation&Ada Lovelace&AD102&161.0&150.1&60.6&85.4&78.7&25.1&46.1&42.2&13.3&18.7&17.4&5.7&6.77&7.85&8.46&6.95  \\
RTX 5070 Ti&Blackwell&GB203&138.7&127.7&47.1&70.7&64.8&19.4&38.9&35.4&9.5&16.3&14.9&4.5&7.93&8.35&9.00&8.59                  \\
RTX PRO 5000 Laptop&Blackwell&GB203&119.1&107.8&34.8&60.1&54.5&15.7&32.9&29.1&8.0&13.5&11.9&3.3&9.49&9.32&11.55&11.85        \\
RTX PRO 5000 48 GB&Blackwell&GB202&154.4&143.0&56.5&91.3&83.2&24.7&51.7&47.0&13.3&22.5&20.5&5.8&7.38&8.87&9.09&8.89          \\
RTX PRO 6000&Blackwell&GB202&171.0&158.6&70.4&107.0&99.8&32.5&62.6&57.9&17.8&27.5&24.6&7.1&7.25&6.73&7.51&10.55              \\
\midrule
\multicolumn{19}{c}{\textbf{Decoding Throughput}}\\\midrule
RTX 2080 SUPER&Turing&TU104&62.9&61.0&24.1&30.0&29.0&11.0&16.8&16.3&6.2&7.6&7.4&2.8&3.02&3.33&2.98&2.63                       \\
RTX 3060 12 GB&Ampere&GA104&42.0&40.5&17.2&20.7&20.0&7.8&11.9&11.6&4.5&5.4&5.3&2.0&3.57&3.38&2.52&1.85                        \\
RTX 3080 12 GB&Ampere&GA102&93.1&89.7&37.2&43.8&41.9&16.4&26.7&25.9&9.9&11.9&11.6&4.4&3.65&4.34&3.00&2.52                     \\
RTX 3090&Ampere&GA102&103.3&99.2&40.9&49.1&47.2&19.3&28.9&28.0&11.4&13.6&13.2&5.1&3.97&3.87&3.11&2.94                         \\
RTX 4060 Ti 16 GB&Ada Lovelace&AD106&68.4&65.5&19.4&29.0&27.9&9.2&15.5&14.9&5.0&6.4&6.2&2.5&4.24&3.79&3.87&3.13               \\
RTX 4070 Ti&Ada Lovelace&AD104&117.7&112.9&36.4&49.8&48.1&16.1&27.1&26.2&9.0&11.1&10.8&4.3&4.08&3.41&3.32&2.70                \\
RTX 4070 Ti SUPER&Ada Lovelace&AD103&122.1&117.5&38.3&57.1&55.0&18.8&31.2&30.1&10.3&13.2&12.8&5.0&3.77&3.68&3.53&3.03         \\
RTX 6000 Ada Generation&Ada Lovelace&AD102&188.5&180.3&49.3&77.8&74.6&22.1&41.5&39.1&10.5&16.7&16.0&4.7&4.35&4.11&5.78&4.19   \\
RTX 5070 Ti&Blackwell&GB203&143.1&136.3&42.2&69.8&66.1&17.2&38.0&35.7&8.2&15.7&14.8&3.8&4.75&5.30&6.05&5.73                   \\
RTX PRO 5000 Laptop&Blackwell&GB203&115.9&117.4&30.8&62.3&58.0&12.5&34.7&32.2&6.5&14.1&13.4&3.1&-1.29&6.90&7.20&4.96          \\
RTX PRO 5000 48 GB&Blackwell&GB202&174.0&165.2&50.9&93.5&87.6&21.6&54.7&51.2&11.5&24.3&22.5&4.8&5.06&6.31&6.40&7.41           \\
RTX PRO 6000&Blackwell&GB202&202.2&191.3&64.8&116.8&109.0&28.7&71.8&67.2&15.1&30.9&28.5&5.6&5.39&6.68&6.41&7.77               \\

\bottomrule
\end{tabular}}
\vspace{-3mm}
\end{table*}

\begin{table*}[!tbp]
\setstretch{0.5}
\caption{PSNR (dB) of the Decoded Output Across Configurations. Compatible Combinations are Highlighted in Blue.}
\vspace{-2mm}
\setlength{\tabcolsep}{1.5pt}
\centering
\label{tab:decodability}
\resizebox{18cm}{!}{\begin{tabular}{@{}lcccccccccccccccccccccccccccccccccccc@{}}
\toprule
\multirow{2.5}{*}{GPU}&\multicolumn{3}{c}{2080S} & \multicolumn{3}{c}{3060} & \multicolumn{3}{c}{3080} & \multicolumn{3}{c}{3090} & \multicolumn{3}{c}{4060 Ti} & \multicolumn{3}{c}{4070 Ti} & \multicolumn{3}{c}{4070 Ti S} & \multicolumn{3}{c}{6000 Ada} & \multicolumn{3}{c}{5070 Ti} & \multicolumn{3}{c}{Pro 5000 LT} & \multicolumn{3}{c}{Pro 5000} & \multicolumn{3}{c}{Pro 6000} \\
\cmidrule(lr){2-4} \cmidrule(lr){5-7} \cmidrule(lr){8-10} \cmidrule(lr){11-13} \cmidrule(lr){14-16} \cmidrule(lr){17-19} \cmidrule(lr){20-22} \cmidrule(lr){23-25}  \cmidrule(lr){26-28}  \cmidrule(lr){29-31}  \cmidrule(lr){32-34}  \cmidrule(lr){35-37}  
&F16&Mix&F32&F16&Mix&F32&F16&Mix&F32&F16&Mix&F32&F16&Mix&F32&F16&Mix&F32&F16&Mix&F32&F16&Mix&F32&F16&Mix&F32&F16&Mix&F32&F16&Mix&F32&F16&Mix&F32
\\

\midrule

\multicolumn{37}{c}{\textbf{1080p/FHD}}\\\midrule

2080S&\textcolor{blue}{\textbf{26.3}}&\textcolor{blue}{\textbf{26.3}}&\textcolor{blue}{\textbf{26.3}}&12.6&13.1&12.7&12.6&13.1&12.7&12.6&13.1&12.7&12.7&13.0&12.7&12.7&13.0&12.7&12.7&13.0&12.7&12.1&12.8&12.7&12.7&12.9&12.7&12.7&12.9&12.7&12.7&12.9&12.7&12.0&12.9&12.7       \\
3060&12.7&11.9&12.8&\textcolor{blue}{\textbf{26.3}}&\textcolor{blue}{\textbf{25.7}}&\textcolor{blue}{\textbf{26.3}}&13.0&\textcolor{blue}{\textbf{26.3}}&\textcolor{blue}{\textbf{26.3}}&13.0&\textcolor{blue}{\textbf{26.3}}&\textcolor{blue}{\textbf{26.3}}&13.2&13.6&\textcolor{blue}{\textbf{26.3}}&13.2&13.6&\textcolor{blue}{\textbf{26.3}}&13.2&13.6&\textcolor{blue}{\textbf{26.3}}&12.9&13.1&\textcolor{blue}{\textbf{26.3}}&11.9&13.6&\textcolor{blue}{\textbf{26.3}}&11.9&13.6&\textcolor{blue}{\textbf{26.3}}&11.9&13.6&\textcolor{blue}{\textbf{26.3}}&12.6&13.6&\textcolor{blue}{\textbf{26.3}}        \\
3080&12.9&11.9&12.9&13.0&\textcolor{blue}{\textbf{26.3}}&\textcolor{blue}{\textbf{26.3}}&\textcolor{blue}{\textbf{26.3}}&\textcolor{blue}{\textbf{26.3}}&\textcolor{blue}{\textbf{26.3}}&\textcolor{blue}{\textbf{26.3}}&\textcolor{blue}{\textbf{26.3}}&\textcolor{blue}{\textbf{26.3}}&13.1&13.6&\textcolor{blue}{\textbf{26.3}}&13.1&13.6&\textcolor{blue}{\textbf{26.3}}&13.1&13.6&\textcolor{blue}{\textbf{26.3}}&13.5&13.1&\textcolor{blue}{\textbf{26.3}}&12.8&13.6&\textcolor{blue}{\textbf{26.3}}&12.8&13.6&\textcolor{blue}{\textbf{26.3}}&12.8&13.6&\textcolor{blue}{\textbf{26.3}}&12.8&13.6&\textcolor{blue}{\textbf{26.3}}        \\
3090&12.9&11.9&12.9&13.0&\textcolor{blue}{\textbf{26.3}}&\textcolor{blue}{\textbf{26.3}}&\textcolor{blue}{\textbf{26.3}}&\textcolor{blue}{\textbf{26.3}}&\textcolor{blue}{\textbf{26.3}}&\textcolor{blue}{\textbf{26.3}}&\textcolor{blue}{\textbf{26.3}}&\textcolor{blue}{\textbf{26.3}}&13.1&13.6&\textcolor{blue}{\textbf{26.3}}&13.1&13.6&\textcolor{blue}{\textbf{26.3}}&13.1&13.6&\textcolor{blue}{\textbf{26.3}}&13.5&13.1&\textcolor{blue}{\textbf{26.3}}&12.8&13.6&\textcolor{blue}{\textbf{26.3}}&12.8&13.6&\textcolor{blue}{\textbf{26.3}}&12.8&13.6&\textcolor{blue}{\textbf{26.3}}&12.8&13.6&\textcolor{blue}{\textbf{26.3}}        \\
4060 Ti&12.9&13.0&12.8&12.6&13.6&\textcolor{blue}{\textbf{26.3}}&13.3&13.6&\textcolor{blue}{\textbf{26.3}}&13.3&13.6&\textcolor{blue}{\textbf{26.3}}&\textcolor{blue}{\textbf{26.3}}&\textcolor{blue}{\textbf{26.3}}&\textcolor{blue}{\textbf{26.3}}&\textcolor{blue}{\textbf{26.3}}&\textcolor{blue}{\textbf{26.3}}&\textcolor{blue}{\textbf{26.3}}&\textcolor{blue}{\textbf{26.3}}&\textcolor{blue}{\textbf{26.3}}&\textcolor{blue}{\textbf{26.3}}&12.9&13.5&\textcolor{blue}{\textbf{26.3}}&12.8&13.5&\textcolor{blue}{\textbf{26.3}}&12.8&13.5&\textcolor{blue}{\textbf{26.3}}&12.8&13.5&\textcolor{blue}{\textbf{26.3}}&12.6&13.5&\textcolor{blue}{\textbf{26.3}}     \\
4070 Ti&12.9&13.0&12.9&12.6&13.6&\textcolor{blue}{\textbf{26.3}}&13.3&13.6&\textcolor{blue}{\textbf{26.3}}&13.3&13.6&\textcolor{blue}{\textbf{26.3}}&\textcolor{blue}{\textbf{26.3}}&\textcolor{blue}{\textbf{26.3}}&\textcolor{blue}{\textbf{26.3}}&\textcolor{blue}{\textbf{26.3}}&\textcolor{blue}{\textbf{26.3}}&\textcolor{blue}{\textbf{26.3}}&\textcolor{blue}{\textbf{26.3}}&\textcolor{blue}{\textbf{26.3}}&\textcolor{blue}{\textbf{26.3}}&12.9&13.5&\textcolor{blue}{\textbf{26.3}}&12.8&13.5&\textcolor{blue}{\textbf{26.3}}&12.8&13.5&\textcolor{blue}{\textbf{26.3}}&12.8&13.5&\textcolor{blue}{\textbf{26.3}}&12.6&13.5&\textcolor{blue}{\textbf{26.3}}     \\
4070 Ti S&12.9&13.0&12.9&12.6&13.6&\textcolor{blue}{\textbf{26.3}}&13.3&13.6&\textcolor{blue}{\textbf{26.3}}&13.3&13.6&\textcolor{blue}{\textbf{26.3}}&\textcolor{blue}{\textbf{26.3}}&\textcolor{blue}{\textbf{26.3}}&\textcolor{blue}{\textbf{26.3}}&\textcolor{blue}{\textbf{26.3}}&\textcolor{blue}{\textbf{26.3}}&\textcolor{blue}{\textbf{26.3}}&\textcolor{blue}{\textbf{26.3}}&\textcolor{blue}{\textbf{26.3}}&\textcolor{blue}{\textbf{26.3}}&12.9&13.5&\textcolor{blue}{\textbf{26.3}}&12.8&13.5&\textcolor{blue}{\textbf{26.3}}&12.8&13.5&\textcolor{blue}{\textbf{26.3}}&12.8&13.5&\textcolor{blue}{\textbf{26.3}}&12.6&13.5&\textcolor{blue}{\textbf{26.3}}   \\
6000 Ada&12.8&13.1&12.9&13.0&13.6&\textcolor{blue}{\textbf{26.3}}&13.1&13.6&\textcolor{blue}{\textbf{26.3}}&13.1&13.6&\textcolor{blue}{\textbf{26.3}}&13.5&13.6&\textcolor{blue}{\textbf{26.3}}&13.5&13.6&\textcolor{blue}{\textbf{26.3}}&13.5&13.6&\textcolor{blue}{\textbf{26.3}}&\textcolor{blue}{\textbf{26.3}}&\textcolor{blue}{\textbf{26.3}}&\textcolor{blue}{\textbf{26.3}}&12.8&13.7&\textcolor{blue}{\textbf{26.3}}&12.8&13.7&\textcolor{blue}{\textbf{26.3}}&12.8&13.7&\textcolor{blue}{\textbf{26.3}}&12.7&13.7&\textcolor{blue}{\textbf{26.3}}    \\
5070 Ti&12.0&12.9&13.0&12.7&13.7&\textcolor{blue}{\textbf{26.3}}&12.7&13.7&\textcolor{blue}{\textbf{26.3}}&12.7&13.7&\textcolor{blue}{\textbf{26.3}}&12.7&13.6&\textcolor{blue}{\textbf{26.3}}&12.7&13.6&\textcolor{blue}{\textbf{26.3}}&12.7&13.6&\textcolor{blue}{\textbf{26.3}}&12.7&13.7&\textcolor{blue}{\textbf{26.3}}&\textcolor{blue}{\textbf{26.3}}&\textcolor{blue}{\textbf{26.3}}&\textcolor{blue}{\textbf{26.3}}&\textcolor{blue}{\textbf{26.3}}&\textcolor{blue}{\textbf{26.3}}&\textcolor{blue}{\textbf{26.3}}&\textcolor{blue}{\textbf{26.3}}&\textcolor{blue}{\textbf{26.3}}&\textcolor{blue}{\textbf{26.3}}&13.0&\textcolor{blue}{\textbf{26.3}}&\textcolor{blue}{\textbf{26.3}}     \\
Pro 5000 LT&12.0&12.9&12.7&12.7&13.6&\textcolor{blue}{\textbf{26.3}}&12.7&13.7&\textcolor{blue}{\textbf{26.3}}&12.7&13.7&\textcolor{blue}{\textbf{26.3}}&12.7&13.6&\textcolor{blue}{\textbf{26.3}}&12.7&13.6&\textcolor{blue}{\textbf{26.3}}&12.7&13.6&\textcolor{blue}{\textbf{26.3}}&12.7&13.7&\textcolor{blue}{\textbf{26.3}}&\textcolor{blue}{\textbf{26.3}}&\textcolor{blue}{\textbf{26.3}}&\textcolor{blue}{\textbf{26.3}}&\textcolor{blue}{\textbf{26.3}}&\textcolor{blue}{\textbf{26.3}}&\textcolor{blue}{\textbf{26.3}}&\textcolor{blue}{\textbf{26.3}}&\textcolor{blue}{\textbf{26.3}}&\textcolor{blue}{\textbf{26.3}}&13.0&\textcolor{blue}{\textbf{26.3}}&\textcolor{blue}{\textbf{26.3}} \\
Pro 5000&12.0&12.9&12.7&12.7&13.7&\textcolor{blue}{\textbf{26.3}}&12.7&13.7&\textcolor{blue}{\textbf{26.3}}&12.7&13.7&\textcolor{blue}{\textbf{26.3}}&12.7&13.6&\textcolor{blue}{\textbf{26.3}}&12.7&13.6&\textcolor{blue}{\textbf{26.3}}&12.7&13.6&\textcolor{blue}{\textbf{26.3}}&12.7&13.7&\textcolor{blue}{\textbf{26.3}}&\textcolor{blue}{\textbf{26.3}}&\textcolor{blue}{\textbf{26.3}}&\textcolor{blue}{\textbf{26.3}}&\textcolor{blue}{\textbf{26.3}}&\textcolor{blue}{\textbf{26.3}}&\textcolor{blue}{\textbf{26.3}}&\textcolor{blue}{\textbf{26.3}}&\textcolor{blue}{\textbf{26.3}}&\textcolor{blue}{\textbf{26.3}}&13.0&\textcolor{blue}{\textbf{26.3}}&\textcolor{blue}{\textbf{26.3}}    \\
Pro 6000&12.3&12.9&12.7&12.7&13.7&\textcolor{blue}{\textbf{26.3}}&12.2&13.7&\textcolor{blue}{\textbf{26.3}}&12.2&13.7&\textcolor{blue}{\textbf{26.3}}&12.6&13.6&\textcolor{blue}{\textbf{26.3}}&12.6&13.6&\textcolor{blue}{\textbf{26.3}}&12.6&13.6&\textcolor{blue}{\textbf{26.3}}&12.2&13.7&\textcolor{blue}{\textbf{26.3}}&13.5&\textcolor{blue}{\textbf{26.3}}&\textcolor{blue}{\textbf{26.3}}&13.5&\textcolor{blue}{\textbf{26.3}}&\textcolor{blue}{\textbf{26.3}}&13.5&\textcolor{blue}{\textbf{26.3}}&\textcolor{blue}{\textbf{26.3}}&\textcolor{blue}{\textbf{26.3}}&\textcolor{blue}{\textbf{26.3}}&\textcolor{blue}{\textbf{26.3}}    \\

\midrule
\multicolumn{37}{c}{\textbf{2160p/4K}}\\\midrule

2080S&\textcolor{blue}{\textbf{28.2}}&\textcolor{blue}{\textbf{28.2}}&\textcolor{blue}{\textbf{28.2}}&7.6&7.5&8.1&7.9&7.5&8.2&7.9&7.5&8.2&7.7&8.1&8.2&7.7&8.1&8.2&7.7&8.1&8.2&8.2&8.1&8.2&7.3&8.1&9.0&7.3&8.1&9.0&7.3&8.1&9.0&7.8&8.1&9.0                       \\
3060&8.1&7.9&9.1&\textcolor{blue}{\textbf{28.2}}&\textcolor{blue}{\textbf{28.2}}&\textcolor{blue}{\textbf{28.2}}&8.7&\textcolor{blue}{\textbf{28.2}}&\textcolor{blue}{\textbf{28.2}}&8.7&\textcolor{blue}{\textbf{28.2}}&\textcolor{blue}{\textbf{28.2}}&8.3&8.6&\textcolor{blue}{\textbf{28.2}}&8.3&8.6&\textcolor{blue}{\textbf{28.2}}&8.3&8.6&\textcolor{blue}{\textbf{28.2}}&8.4&8.6&\textcolor{blue}{\textbf{28.2}}&8.1&9.2&10.2&8.1&9.2&10.2&8.1&9.2&10.2&8.1&9.2&10.2            \\
3080&7.3&7.9&8.9&8.3&\textcolor{blue}{\textbf{28.2}}&\textcolor{blue}{\textbf{28.2}}&\textcolor{blue}{\textbf{28.2}}&\textcolor{blue}{\textbf{28.2}}&\textcolor{blue}{\textbf{28.2}}&\textcolor{blue}{\textbf{28.2}}&\textcolor{blue}{\textbf{28.2}}&\textcolor{blue}{\textbf{28.2}}&9.0&8.6&\textcolor{blue}{\textbf{28.2}}&9.0&8.6&\textcolor{blue}{\textbf{28.2}}&9.0&8.6&\textcolor{blue}{\textbf{28.2}}&9.1&8.6&\textcolor{blue}{\textbf{28.2}}&8.0&9.2&9.4&8.0&9.2&9.4&8.0&9.2&9.4&8.2&9.2&9.4               \\
3090&7.3&7.9&9.1&8.3&\textcolor{blue}{\textbf{28.2}}&\textcolor{blue}{\textbf{28.2}}&\textcolor{blue}{\textbf{28.2}}&\textcolor{blue}{\textbf{28.2}}&\textcolor{blue}{\textbf{28.2}}&\textcolor{blue}{\textbf{28.2}}&\textcolor{blue}{\textbf{28.2}}&\textcolor{blue}{\textbf{28.2}}&9.0&8.6&\textcolor{blue}{\textbf{28.2}}&9.0&8.6&\textcolor{blue}{\textbf{28.2}}&9.0&8.6&\textcolor{blue}{\textbf{28.2}}&9.1&8.6&\textcolor{blue}{\textbf{28.2}}&8.0&9.2&10.4&8.0&9.2&10.4&8.0&9.2&10.4&8.2&9.2&10.4           \\
4060 Ti&7.9&7.8&9.1&8.4&8.9&\textcolor{blue}{\textbf{28.2}}&8.4&8.9&\textcolor{blue}{\textbf{28.2}}&8.4&8.9&\textcolor{blue}{\textbf{28.2}}&\textcolor{blue}{\textbf{28.2}}&\textcolor{blue}{\textbf{28.2}}&\textcolor{blue}{\textbf{28.2}}&\textcolor{blue}{\textbf{28.2}}&\textcolor{blue}{\textbf{28.2}}&\textcolor{blue}{\textbf{28.2}}&\textcolor{blue}{\textbf{28.2}}&\textcolor{blue}{\textbf{28.2}}&\textcolor{blue}{\textbf{28.2}}&9.7&\textcolor{blue}{\textbf{28.2}}&\textcolor{blue}{\textbf{28.2}}&7.7&8.8&10.2&7.7&8.8&10.2&7.7&8.8&10.2&7.7&8.8&10.2      \\
4070 Ti&7.9&7.8&8.8&8.4&8.9&\textcolor{blue}{\textbf{28.2}}&8.4&8.9&\textcolor{blue}{\textbf{28.2}}&8.4&8.9&\textcolor{blue}{\textbf{28.2}}&\textcolor{blue}{\textbf{28.2}}&\textcolor{blue}{\textbf{28.2}}&\textcolor{blue}{\textbf{28.2}}&\textcolor{blue}{\textbf{28.2}}&\textcolor{blue}{\textbf{28.2}}&\textcolor{blue}{\textbf{28.2}}&\textcolor{blue}{\textbf{28.2}}&\textcolor{blue}{\textbf{28.2}}&\textcolor{blue}{\textbf{28.2}}&9.7&\textcolor{blue}{\textbf{28.2}}&\textcolor{blue}{\textbf{28.2}}&7.7&8.8&10.6&7.7&8.8&10.6&7.7&8.8&10.6&7.7&8.8&10.6      \\
4070 Ti S&7.9&7.8&8.8&8.4&8.9&\textcolor{blue}{\textbf{28.2}}&8.4&8.9&\textcolor{blue}{\textbf{28.2}}&8.4&8.9&\textcolor{blue}{\textbf{28.2}}&\textcolor{blue}{\textbf{28.2}}&\textcolor{blue}{\textbf{28.2}}&\textcolor{blue}{\textbf{28.2}}&\textcolor{blue}{\textbf{28.2}}&\textcolor{blue}{\textbf{28.2}}&\textcolor{blue}{\textbf{28.2}}&\textcolor{blue}{\textbf{28.2}}&\textcolor{blue}{\textbf{28.2}}&\textcolor{blue}{\textbf{28.2}}&9.7&\textcolor{blue}{\textbf{28.2}}&\textcolor{blue}{\textbf{28.2}}&7.7&8.8&10.6&7.7&8.8&10.6&7.7&8.8&10.6&7.7&8.8&10.6    \\
6000 Ada&8.0&7.8&9.1&9.1&8.9&\textcolor{blue}{\textbf{28.2}}&8.7&8.9&\textcolor{blue}{\textbf{28.2}}&8.7&8.9&\textcolor{blue}{\textbf{28.2}}&9.5&\textcolor{blue}{\textbf{28.2}}&\textcolor{blue}{\textbf{28.2}}&9.5&\textcolor{blue}{\textbf{28.2}}&\textcolor{blue}{\textbf{28.2}}&9.5&\textcolor{blue}{\textbf{28.2}}&\textcolor{blue}{\textbf{28.2}}&\textcolor{blue}{\textbf{28.2}}&\textcolor{blue}{\textbf{28.2}}&\textcolor{blue}{\textbf{28.2}}&7.9&8.8&10.4&7.9&8.8&10.4&7.9&8.8&10.4&7.5&8.8&10.4       \\
5070 Ti&8.0&8.0&9.0&8.0&9.1&9.9&8.2&9.1&9.9&8.2&9.1&9.9&8.1&9.1&9.9&8.1&9.1&9.9&8.1&9.1&9.9&8.1&9.1&9.9&\textcolor{blue}{\textbf{28.2}}&\textcolor{blue}{\textbf{28.2}}&\textcolor{blue}{\textbf{28.2}}&\textcolor{blue}{\textbf{28.2}}&\textcolor{blue}{\textbf{28.2}}&\textcolor{blue}{\textbf{28.2}}&\textcolor{blue}{\textbf{28.2}}&\textcolor{blue}{\textbf{28.2}}&\textcolor{blue}{\textbf{28.2}}&8.8&\textcolor{blue}{\textbf{28.2}}&\textcolor{blue}{\textbf{28.2}}             \\
Pro 5000 LT&8.0&8.0&8.5&8.0&9.1&10.2&8.2&9.1&10.2&8.2&9.1&10.2&8.1&9.1&10.2&8.1&9.1&10.2&8.1&9.1&10.2&8.1&9.1&10.2&\textcolor{blue}{\textbf{28.2}}&\textcolor{blue}{\textbf{28.2}}&\textcolor{blue}{\textbf{28.2}}&\textcolor{blue}{\textbf{28.2}}&\textcolor{blue}{\textbf{28.2}}&\textcolor{blue}{\textbf{28.2}}&\textcolor{blue}{\textbf{28.2}}&\textcolor{blue}{\textbf{28.2}}&\textcolor{blue}{\textbf{28.2}}&8.8&\textcolor{blue}{\textbf{28.2}}&\textcolor{blue}{\textbf{28.2}}  \\
Pro 5000&8.0&8.0&8.5&8.0&9.1&10.2&8.2&9.1&10.2&8.2&9.1&10.2&8.1&9.1&10.2&8.1&9.1&10.2&8.1&9.1&10.2&8.1&9.1&10.2&\textcolor{blue}{\textbf{28.2}}&\textcolor{blue}{\textbf{28.2}}&\textcolor{blue}{\textbf{28.2}}&\textcolor{blue}{\textbf{28.2}}&\textcolor{blue}{\textbf{28.2}}&\textcolor{blue}{\textbf{28.2}}&\textcolor{blue}{\textbf{28.2}}&\textcolor{blue}{\textbf{28.2}}&\textcolor{blue}{\textbf{28.2}}&8.8&\textcolor{blue}{\textbf{28.2}}&\textcolor{blue}{\textbf{28.2}}     \\
Pro 6000&7.9&8.0&9.3&7.6&9.1&9.9&8.2&9.1&9.9&8.2&9.1&9.9&8.2&9.1&9.9&8.2&9.1&9.9&8.2&9.1&9.9&8.3&9.1&9.9&8.4&\textcolor{blue}{\textbf{28.2}}&\textcolor{blue}{\textbf{28.2}}&8.4&\textcolor{blue}{\textbf{28.2}}&\textcolor{blue}{\textbf{28.2}}&8.4&\textcolor{blue}{\textbf{28.2}}&\textcolor{blue}{\textbf{28.2}}&\textcolor{blue}{\textbf{28.2}}&\textcolor{blue}{\textbf{28.2}}&\textcolor{blue}{\textbf{28.2}}              \\

\bottomrule
\end{tabular}}
\vspace{-7mm}
\end{table*}

\subsection{Coding Throughput} \label{sec:coding_thpt}

Table \ref{tab:Thpt} presents encoding and decoding throughput across all tested GPUs, resolutions, and precision modes. Mixed precision incurs a modest throughput penalty relative to full FP16, averaging $\approx6.8\%$ for encoding and $\approx4.2\%$ for decoding across all resolutions. The encoding penalty is consistently higher because the FP32 I-net is computationally heavier to encode than to decode. Decoding is slower than encoding on most GPUs, whereas the fastest GPUs (e.g., RTX PRO 6000) decode faster than they encode. Overall, the mixed precision overhead is modest and consistent, remaining below 12\% in all configurations, which we consider acceptable given the cross-GPU determinism benefits. Full FP32 inference imposes a substantial throughput reduction relative to FP16, approximately 3-4× across all GPUs and resolutions. This is consistent with the theoretical throughput ratio between CUDA core FP32 and tensor core FP16 operations on modern NVIDIA GPUs. Beyond the throughput difference, FP32 is also more power-intensive. GPU computational performance is proportional to its clock speed: as a GPU hits the power limits, it throttles the shader clock significantly, reducing computational throughput. Tensor core FP16 execution is more energy-efficient per operation, which leaves more power headroom, resulting in higher sustained clock speed.

\begin{table}[!tbp]

\setstretch{0.70}
\vspace{1mm}
\caption{End-to-End Latency  (ms)}
\vspace{-2mm}
\centering
\label{tab:E2ELatency}
\resizebox{8.5cm}{!}{\begin{tabular}{@{}lccccccc@{}}
\toprule
\multirow{2.5}{*}{Network Type}& \multicolumn{2}{c}{Ping (ICMP)} & \multicolumn{5}{c}{Live Encoder/Decoder} \\
\cmidrule(lr){2-3} \cmidrule(lr){4-8}
& Avg. & mdev & I Avg. & P Avg. & Avg. & Med & 95\%-ile\\
\midrule
Ethernet&7.65&0.67&116.71&41.15&41.78&36.84&62.87        \\
Wi-Fi&11.85&1.57&149.88&55.17&55.96&44.91&94.26            \\
FDD-LTE&30.33&3.63&158.00&66.91&67.67&60.33&89.02            \\
TDD-LTE&32.83&5.34&144.95&61.15&61.85&54.11&85.75            \\
FDD-NR&27.91&4.59&138.57&58.97&59.64&53.12&80.84            \\
FDD-NR (DSS)&28.96&5.93&146.90&60.75&61.47&52.75&82.61    \\
TDD-NR&35.55&7.84&140.62&59.53&60.21&53.52&82.87            \\
Starlink D2C&\textcolor{red}{\textbf{188.57}}&\textcolor{red}{\textbf{98.31}}&\textcolor{red}{\textbf{1376.96}}&\textcolor{red}{\textbf{1272.07}}&\textcolor{red}{\textbf{1272.95}}&\textcolor{red}{\textbf{426.89}}&\textcolor{red}{\textbf{5583.25}}  \\
\bottomrule
\end{tabular}}
\vspace{-8mm}
\end{table}

This power sensitivity manifests clearly in the comparison between the RTX 3090 and the RTX PRO 5000 Laptop. The RTX 3090, with a 390W power budget, sustains higher shader clocks under load and outperforms the PRO 5000 Laptop in FP32 throughput at all resolutions (e.g., 21.6 vs. 15.7 fps encoding at 1080p FP32). Under FP16, however, the balance reverses as the power limit is no longer an issue. This allows the PRO 5000 Laptop to sustain high boost clock speeds, achieving 60.10 fps encoding throughput at 1080p FP16 compared to 51.50 fps on the RTX 3090. Consequently, the mixed precision performance penalty relative to FP16 varies across GPUs depending on their power envelope: GPUs with generous power budgets show a smaller relative penalty because their FP32 I-frame performance is less throttled, while power-constrained GPUs show a larger relative gap between FP16 and FP32, making mixed precision a more attractive compromise. Moreover, at lower resolutions, throughput is primarily compute-bound, where execution units are the limiting resource. At higher resolutions, the feature maps grow large and require repeated VRAM accesses, which makes the memory bandwidth the dominant bottleneck. This transition produces a resolution-dependent reversal in the relative ranking of some GPU pairs. The RTX 4060 Ti (AD106, 288 GB/s memory bandwidth, 4352 CUDA cores) outperforms the RTX 2080 SUPER (TU104, 496 GB/s memory bandwidth, 3072 CUDA cores) at 720p in FP16 and Mixed Precision, because at that resolution the workload is compute-bound and the 4060 Ti's higher core count and newer tensor core generation are the decisive factors. At 1080p and above the advantage shifts to the RTX 2080 SUPER, whose nearly 1.7× memory bandwidth advantage compensates for its older architecture as the workload transitions toward memory-bound execution. This finding suggests that memory bandwidth should be weighted alongside compute TFLOPS when selecting hardware for neural video codec deployment, particularly for high-resolution applications.

\vspace{-1.2mm}

\subsection{Cross-Decodability}\label{sec:cross_dec}

To evaluate hardware interoperability, we analyzed the bitstreams generated across all tested GPUs. We found that the baseline FP16 mode produces identical bitstreams (matching SHA-256 hashes) within the same GPU generation, with the exception of the highest-SM-count GPUs in each generation, which generate different bitstreams. The proposed Mixed Precision mode resolves this intra-generation fragmentation, yielding bit-exact identical hashes across all die variants within the same generation at 4K (at 1080p, the RTX 6000 Ada remains the only exception) and significantly expanding cross-decodability. Since the P-net runs in FP16 in both modes, the die-dependent computations of the FP16 mode must lie in the I-frame path. For both FP16 and Mixed Precision, whenever a GPU successfully decodes a bitstream, the resulting decoded output is identical across devices. The full FP32 mode exhibits a fundamentally different behavior. While it is strictly non-deterministic, with each GPU architecture generating a unique bitstream with a different SHA-256 hash, it remains highly inter-decodable. Furthermore, when fed the same encoded bitstream, all GPUs produce perfectly identical decoded outputs. Interestingly, we discovered that FP32 cross-decodability is highly resolution-dependent (as reflected in Table \ref{tab:decodability}). At 1080p, FP32 bitstreams are seamlessly inter-decodable across all Ampere, Ada Lovelace, and Blackwell GPUs. This broad, three-generation cross-decodability holds consistently for resolutions between 1760×990 and 3520×1980. However, at 4K (2160p), decodability fractures into three isolated compatibility groups: (1) Turing, (2) Ampere and Ada Lovelace, and (3) Blackwell. Finally, the Turing architecture remains universally incompatible with newer hardware. Thus, higher precision naturally widens decodability even when strict determinism fails. We hypothesize that the 4K fracture stems from resolution-dependent kernel selection: at 4K, tensor shapes cross cuDNN heuristic thresholds and dispatch generation-specific kernels whose differing reduction orders break bit-exactness even under IEEE-754 FP32, which fixes per-operation rounding but not operation ordering. Kernel-level tracing to validate this is left for future work.

\vspace{-1.5mm}

\subsection{Compression Efficiency}

To verify that the proposed precision modes do not degrade rate-distortion performance, we evaluated the compression efficiency across all tested GPUs and resolutions. Extensive testing across all 12 GPUs at 1080p and 2160p revealed average BD-Rate deviations of at most +0.0162\% (Mixed) and -0.0112\% (FP32), with no individual GPU exceeding 0.04\%, confirming virtually identical compression efficiency. These fluctuations fall well below the 0.1\% threshold generally considered to be measurement noise in video coding evaluations. Consequently, the only cost incurred by utilizing Mixed Precision or FP32 is the reduction in coding throughput discussed in Section \ref{sec:coding_thpt}, while the compression efficiency is unaffected.

\vspace{-1.2mm}

\subsection{End-to-End Latency}

Table \ref{tab:E2ELatency} summarizes the end-to-end round-trip latency across the evaluated network types. Crucially, our results demonstrate that the proposed streamable LVC architecture successfully operates across all tested environments, from ideal wired connections to highly variable Low Earth Orbit (LEO) satellite links. As expected, wired Ethernet provides the most stable baseline, exhibiting the lowest average loaded latency (41.78 ms) and minimal jitter (62.87 ms at the 95th percentile). While Wi-Fi 6 yields significantly faster unloaded latency (11.85 ms ICMP ping) compared to its cellular counterparts, its loaded latency (55.96 ms average) degrades to levels comparable to optimal cellular conditions, reflecting the impact of wireless medium contention during sustained, high-throughput transmission.

Within the terrestrial cellular configurations, a distinct dynamic emerges between Frequency Division Duplexing (FDD) and Time Division Duplexing (TDD). Across both LTE and 5G NR Radio Access Technologies (RATs), FDD provides strictly lower unloaded latency (e.g., 27.91 ms for FDD-NR vs. 35.55 ms for TDD-NR). However, under the DCVC-RT streaming load, this advantage narrows or inverts. The wider downlink allocations typical of TDD configurations significantly accelerate network transit times for large video packets: the loaded latency shifts in TDD's favor on LTE (61.85 ms vs. 67.67 ms for FDD-LTE), while the gap closes to 0.6 ms on NR. Additionally, we observe a distinct performance penalty when Dynamic Spectrum Sharing (DSS) is enabled on FDD-NR; the DSS overhead negatively impacts both average loaded latency and 95th percentile jitter compared to dedicated FDD-NR spectrum. \looseness=-5

Finally, we evaluated the system over a Non-Terrestrial Network (NTN) using Starlink D2C. From the perspective of the User Equipment (UE), Starlink D2C operates identically to a terrestrial base station, utilizing standard LTE signaling over a 5 MHz channel (Band 1). While our architecture successfully sustained the decoded stream, the network exhibited severe delay and jitter, with loaded latency averaging 1272.95 ms and a 95th percentile exceeding 5.5 seconds. Our analysis found that this is due to satellite beam handovers, which take approximately 3 to 6 seconds to complete---far slower than the millisecond-level handovers of terrestrial networks. This prolonged handover duration causes severe bufferbloat. Furthermore, available throughput drops precipitously from a peak of 5 Mbps to mere kilobits per second as the UE transitions toward the edge of a beam's coverage area. These extreme network dynamics highlight a critical direction for future work: developing robust, highly rate-adaptive, and error-resilient learned video codecs tailored specifically to the high-jitter, variable-throughput characteristics of emerging NTNs. \looseness=-5

\section{Conclusion and Future Work}

Deploying learned video compression in real-world streaming environments requires overcoming strict constraints regarding cross-hardware determinism, real-time throughput, and network volatility. In this paper, we introduced a streamable, client-server NVC architecture that resolves floating-point non-determinism without resorting to the computationally crippling or efficiency-destroying compromises of integer quantization. By employing a novel Mixed Precision strategy alongside deterministic DPB state alignment, we leverage FP16 Tensor Cores for high-speed predictive coding while utilizing FP32 CUDA cores for the die-sensitive computations at critical GOP boundaries.

Extensive hardware evaluations across 12 distinct GPUs revealed that our approach largely eliminates intra-generation bitstream fragmentation and significantly expands inter-decodability across multiple architectures, achieving broad cross-generation compatibility at 1080p resolution. We demonstrated that this interoperability enhancement is achieved at virtually zero cost to compression efficiency, with BD-Rate PSNR deviations remaining below 0.04\%. Furthermore, end-to-end latency evaluations over operational networks confirmed the system's viability across Ethernet, Wi-Fi 6, and 5G NR, while revealing load-dependent latency shifts between TDD and FDD cellular configurations.

While our architecture successfully sustained decoded streams over Starlink Direct-to-Cell, the extreme multi-second jitter introduced by satellite beam handovers highlights a vital area for future research. Future work will focus on developing error-resilient, highly rate-adaptive NVC mechanisms tailored specifically to the extreme dynamics of Non-Terrestrial Networks (NTNs). Additionally, future investigations will validate the kernel-selection hypothesis behind the 4K interoperability fractures via kernel-level tracing. Finally, the source code remains under active development; extensive validation is currently underway to address these remaining edge cases prior to a public open-source release.

\vspace{-1mm}
\section*{Acknowledgement}

This paper is supported by the Ministry of Internal Affairs and Communications (MIC) Project for Efficient Frequency Utilization Toward Wireless IP Multicasting and the Japan Science and Technology Agency (JST) CRONOS Grant Number JPMJCS25N2. Additionally, the authors thank \textit{Qtrun Technologies} for providing \textit{Network Signal Guru (NSG)}.




%
\setstretch{0.81}
\Urlmuskip=0mu plus 1mu\relax
\bibliographystyle{IEEEtran}
\bibliography{IEEEabrv,b_reference}

@IEEEtranBSTCTL{IEEEexample:BSTcontrol,
  CTLuse_forced_etal       = "yes",
  CTLmax_names_forced_etal = "2",
  CTLnames_show_etal       = "2"
}

@INPROCEEDINGS{11396901,
  author={Arunruangsirilert, Kasidis and Katto, Jiro},
  booktitle={2025 International Conference on Visual Communications and Image Processing (VCIP)}, 
  title={Evaluation of GPU Video Encoder for Low-Latency Real-Time 4K UHD Encoding}, 
  year={2025},
  volume={},
  number={},
  pages={1-5},
  doi={10.1109/VCIP67698.2025.11396901}}

@misc{barman2024bjontegaarddeltabdtutorial,
      title={Bj{\o}ntegaard Delta (BD): A Tutorial Overview of the Metric, Evolution, Challenges, and Recommendations}, 
      author={Nabajeet Barman and Maria G. Martini and Yuriy Reznik},
      year={2024},
      eprint={2401.04039},
      archivePrefix={arXiv},
      primaryClass={cs.MM},
      url={https://arxiv.org/abs/2401.04039}, 
}

@INPROCEEDINGS{11417632,
  author={Arunruangsirilert, Kasidis and Katto, Jiro},
  booktitle={2025 Picture Coding Symposium (PCS)}, 
  title={Evaluation of NVENC Split-Frame Encoding (SFE) for UHD Video Transcoding}, 
  year={2025},
  volume={},
  number={},
  pages={1-5},
  doi={10.1109/PCS65673.2025.11417632}}

@misc{tian2023effortlesscrossplatformvideocodec,
      title={Effortless Cross-Platform Video Codec: A Codebook-Based Method}, 
      author={Kuan Tian and Yonghang Guan and Jinxi Xiang and Jun Zhang and Xiao Han and Wei Yang},
      year={2023},
      eprint={2310.10292},
      archivePrefix={arXiv},
      primaryClass={cs.CV},
      url={https://arxiv.org/abs/2310.10292}, 
}

@INPROCEEDINGS{10448359,
  author={Koyuncu, Esin and Solovyev, Timofey and Sauer, Johannes and Alshina, Elena and Kaup, André},
  booktitle={ICASSP 2024 - 2024 IEEE International Conference on Acoustics, Speech and Signal Processing (ICASSP)}, 
  title={Quantized Decoder in Learned Image Compression for Deterministic Reconstruction}, 
  year={2024},
  volume={},
  number={},
  pages={3985-3989},
  doi={10.1109/ICASSP48485.2024.10448359}}

@misc{burns_2024, title={A guide to AI TOPS and NPU performance metrics}, url={https://www.qualcomm.com/news/onq/2024/04/a-guide-to-ai-tops-and-npu-performance-metrics}, journal={Qualcomm.com}, author={Burns, Peter}, year={2024} }

@misc{nvidia_tensorcore, title={NVIDIA Tensor Cores: Versatility for HPC \& AI}, url={https://www.nvidia.com/en-us/data-center/tensor-cores/}, journal={NVIDIA}, author={NVIDIA} }

@INPROCEEDINGS{10484439,
  author={van Rozendaal, Ties and Singhal, Tushar and Le, Hoang and Sautiere, Guillaume and Said, Amir and Buska, Krishna and Raha, Anjuman and Kalatzis, Dimitris and Mehta, Hitarth and Mayer, Frank and Zhang, Liang and Nagel, Markus and Wiggers, Auke},
  booktitle={2024 IEEE/CVF Winter Conference on Applications of Computer Vision (WACV)}, 
  title={MobileNVC: Real-time 1080p Neural Video Compression on a Mobile Device}, 
  year={2024},
  volume={},
  number={},
  pages={4311-4321},
  doi={10.1109/WACV57701.2024.00427}}

@misc{
jia2026integercentric,
title={Integer-Centric Neural Video Compression},
author={Zhaoyang Jia and Wenxuan Xie and Zongyu Guo and Bin Li and Jiahao Li and Houqiang Li and Yan Lu},
year={2026},
url={https://openreview.net/forum?id=KCQo0fXtFH}
}

@inproceedings{10.1145/3581783.3611955,
author = {Tian, Kuan and Guan, Yonghang and Xiang, Jinxi and Zhang, Jun and Han, Xiao and Yang, Wei},
title = {Towards Real-Time Neural Video Codec for Cross-Platform Application Using Calibration Information},
year = {2023},
isbn = {9798400701085},
publisher = {Association for Computing Machinery},
address = {New York, NY, USA},
url = {https://doi.org/10.1145/3581783.3611955},
doi = {10.1145/3581783.3611955},
booktitle = {Proceedings of the 31st ACM International Conference on Multimedia},
pages = {7961–7970},
numpages = {10},
location = {Ottawa ON, Canada},
series = {MM '23}
}

@INPROCEEDINGS{10743422,
  author={Pang, Jiahao and Lodhi, Muhammad Asad and Ahn, Junghyun and Huang, Yuning and Tian, Dong},
  booktitle={2024 IEEE 26th International Workshop on Multimedia Signal Processing (MMSP)}, 
  title={Towards Reproducible Learning-Based Compression}, 
  year={2024},
  volume={},
  number={},
  pages={1-6},
  doi={10.1109/MMSP61759.2024.10743422}}

@ARTICLE{10089871,
  author={Danesh Pazho, Armin and Neff, Christopher and Noghre, Ghazal Alinezhad and Ardabili, Babak Rahimi and Yao, Shanle and Baharani, Mohammadreza and Tabkhi, Hamed},
  journal={IEEE Internet of Things Journal}, 
  title={Ancilia: Scalable Intelligent Video Surveillance for the Artificial Intelligence of Things}, 
  year={2023},
  volume={10},
  number={17},
  pages={14940-14951},
  doi={10.1109/JIOT.2023.3263725}}

@ARTICLE{10731639,
  author={Sehad, Nassim and Bariah, Lina and Hamidouche, Wassim and Hellaoui, Hamed and Jäntti, Riku and Debbah, Mérouane},
  journal={IEEE Communications Magazine}, 
  title={Generative AI for Immersive Communication: The Next Frontier in Internet-of-Senses Through 6G}, 
  year={2025},
  volume={63},
  number={2},
  pages={31-43},
  doi={10.1109/MCOM.001.2400199}}

@ARTICLE{11016906,
  author={Trigka, Maria and Dritsas, Elias},
  journal={IEEE Access}, 
  title={The Evolution of Generative AI: Trends and Applications}, 
  year={2025},
  volume={13},
  number={},
  pages={98504-98529},
  doi={10.1109/ACCESS.2025.3574660}}

@ARTICLE{4801529,
  author={Cha, Meeyoung and Kwak, Haewoon and Rodriguez, Pablo and Ahn, Yong-Yeol and Moon, Sue},
  journal={IEEE/ACM Transactions on Networking}, 
  title={Analyzing the Video Popularity Characteristics of Large-Scale User Generated Content Systems}, 
  year={2009},
  volume={17},
  number={5},
  pages={1357-1370},
  doi={10.1109/TNET.2008.2011358}}

@INPROCEEDINGS{11099929,
  author={Wang, Peng and Zhang, Qinjuan and Ma, Yujuan and Lu, Shan and Deng, Yao and Duo, Hao},
  booktitle={2025 4th International Conference on Electronics, Integrated Circuits and Communication Technology (EICCT)}, 
  title={Status and Development Analysis of Satellite-Direct-to-Phone Technology for Existing Handsets}, 
  year={2025},
  volume={},
  number={},
  pages={295-299},
  doi={10.1109/EICCT65471.2025.11099929}}

@INPROCEEDINGS{11133825,
  author={Arunruangsirilert, Kasidis and Wongprasert, Pasapong and Katto, Jiro},
  booktitle={2025 34th International Conference on Computer Communications and Networks (ICCCN)}, 
  title={Evaluations of High Power User Equipment (HPUE) in Urban Environment}, 
  year={2025},
  volume={},
  number={},
  pages={1-6},
  doi={10.1109/ICCCN65249.2025.11133825}}

@INPROCEEDINGS{11095025,
  author={Jia, Zhaoyang and Li, Bin and Li, Jiahao and Xie, Wenxuan and Qi, Linfeng and Li, Houqiang and Lu, Yan},
  booktitle={2025 IEEE/CVF Conference on Computer Vision and Pattern Recognition (CVPR)}, 
  title={Towards Practical Real-Time Neural Video Compression}, 
  year={2025},
  volume={},
  number={},
  pages={12543-12552},
  doi={10.1109/CVPR52734.2025.01170}}

@inproceedings{10.1145/3746027.3755598,
author = {Liao, Junqi and Wu, Yaojun and Lin, Chaoyi and Deng, Zhipin and Li, Li and Liu, Dong and Sun, Xiaoyan},
title = {EHVC: Efficient Hierarchical Reference and Quality Structure for Neural Video Coding},
year = {2025},
isbn = {9798400720352},
publisher = {Association for Computing Machinery},
address = {New York, NY, USA},
url = {https://doi.org/10.1145/3746027.3755598},
doi = {10.1145/3746027.3755598},
booktitle = {Proceedings of the 33rd ACM International Conference on Multimedia},
pages = {12083–12091},
numpages = {9},
location = {Dublin, Ireland},
series = {MM '25}
}

@ARTICLE{9503377,
  author={Bross, Benjamin and Wang, Ye-Kui and Ye, Yan and Liu, Shan and Chen, Jianle and Sullivan, Gary J. and Ohm, Jens-Rainer},
  journal={IEEE Transactions on Circuits and Systems for Video Technology}, 
  title={Overview of the Versatile Video Coding (VVC) Standard and its Applications}, 
  year={2021},
  volume={31},
  number={10},
  pages={3736-3764},
  doi={10.1109/TCSVT.2021.3101953}}

@ARTICLE{6316136,
  author={Sullivan, Gary J. and Ohm, Jens-Rainer and Han, Woo-Jin and Wiegand, Thomas},
  journal={IEEE Transactions on Circuits and Systems for Video Technology}, 
  title={Overview of the High Efficiency Video Coding (HEVC) Standard}, 
  year={2012},
  volume={22},
  number={12},
  pages={1649-1668},
  doi={10.1109/TCSVT.2012.2221191}}

@misc{xiph.org, title={Xiph.org :: Derf’s Test Media Collection}, url={https://media.xiph.org/video/derf/}, journal={media.xiph.org}, author={Xiph.org} }

@misc{ITE_2016, title={Ultra-high definition/wide-color-gamut standard test sequences – Series A}, url={https://www.ite.or.jp/content/test-materials/uhdtv_a/}, journal={Ultra-high definition/wide-color-gamut standard test sequences – Series A}, author={The Institute of Image Information and Television Engineers}, year={2016}, month={Jan} }

@misc{google, title={Choose live encoder settings, bitrates, and resolutions - YouTube Help}, url={https://support.google.com/youtube/answer/2853702?hl=en}, journal={support.google.com}, author={Google} }

@misc{twitch, title={Twitch Help Portal}, url={https://help.twitch.tv/s/article/broadcasting-guidelines?language=en_US}, journal={help.twitch.tv}, author={Twitch} }

\end{document}